\documentclass[12pt]{iopart}
\usepackage[latin1]{inputenc}
\usepackage[final]{graphicx}

\begin{document}

\title{Underlying Events in $pp$ Collisions at LHC Energies}

\author{Andr\'as G. Ag\'ocs$^{1,2}$, Gergely G. Barnaf\"oldi$^1$ and P\'eter L\'evai$^1$}

\address{$^1$KFKI Research Institute for Particle and Nuclear Physics of the HAS\\
29-33 Konkoly-Thege Mikl\'os Str. H-1121 Budapest.  \smallskip\\
$^2$E\"otv\"os University, Faculty of Sciences\\
1/A P\'azm\'any P\'eter s\'et\'any, H-1117 Budapest.}
\ead{agocs@rmki.kfki.hu}

\begin{abstract}
Hadron production is investigated in proton-proton ($pp$) collisions at 
$\sqrt{s}=7$ TeV energy -- especially outside the cones of identified jets. 
We have improved the original CDF definition of Underlying Event and have introduced
surrounding rings/belts (SB) around the cone of identified jets. 
We compare the characteristics of hadron production using the 
CDF-based and the new SB-based methods analyzing PYTHIA generated 
hadron production within different geometrical regions of $pp$ collisions.
\end{abstract}

\section{Introduction}

The study of hadron showers (jets) started at the LEP and TEVATRON experiments 
and have been continued at RHIC (BNL) and LHC (CERN). During the last two decades 
these experiments focused on electron-positron ($e^-e^+$) and proton-(anti)proton 
($pp$, $p\bar{p}$) collisions. 
Various methods of jet-identification have been tested successfully. 

Investigating jets in high energy hadron-hadron collisions can provide
an understanding of the complex physics of strong interaction and even
theories beyond the Standard Model. Recent state-of-the-art analysis
techniques~\cite{Salam:2009,Salam:2010} reached the point to able to
analyze the inner structure of jets, surrounding area of hadron
showers, off-jet directions in nucleon-nucleon, and jet-matter
interaction in nucleus-nucleus collisions. These latter studies are
connected to the research of quark-gluon plasma (QGP), which strongly
requires the \ separation of hard (perturbative) and soft (background)
component of the reactions.

The Underlying Event (UE) was introduced by the CDF Collaboration 
at TEVATRON energies~\cite{CDFUE} to separate and
identify soft (or semi-hard) contributions of a high-energy collision. 
Since multiple jet events were very rare, then UE could represent the remaining 
hadrons of a $p{\bar p}$ collisions, after the leading jet was identified
and extracted. 
The CDF-definition corresponds to jet identification in one-jet events, 
where the second jet is assumed to move automatically into the away side. 
An improved version of the CDF definition can be applied for multi-jet events, 
namely introducing Surrounding Belts (SB) around 
identified jets~\cite{Agocs:2009,Agocs:2010}.

\section{The New Definition of UE and the Concept}

The CDF definition of the underlying event is a simple and practical tool, 
since opens the jet angle acceptances to the maximum: 
$1/3$ to the near,  $1/3$ to the away, and finally $1/3$ for the 
two transverse regimes called the UE. Moreover, the CDF event 
geometry can be fixed easily, since the position of the leading 
jet defines the {\sl toward region}, and the {\sl away region} 
will be chosen respectively~\cite{CDFUE}. Thus, hadrons moving 
to the {\sl transverse directions} are assigned to be off-jet, 
background particles. The weakness of the CDF-based UE definition 
is that it assumes a single or back-to-back jet-event situation, 
which is not always the case at higher energies. In case of multiple 
jets or jet-matter interaction with secondary collisions 
the hadron content of the CDF-regions would mix up.  
The question is: how can we identify and separate the proper regions?

Our idea was to improve and develop a new UE definition, which is 
strongly connected to the identified jets (excluding all charged 
particles from all identified jets), independent of the number of jets. 
Moreover, jet-matter-interaction secondaries can be also separated and 
investigated within the surrounding areas around identified jets. 
By this method the study of $pp$ and $AA$ collisions can be done 
in the same framework without major changes in jet-finding parameters.

\vspace{-1.0truecm}
\begin{figure*}[ht]
  \centering
  \begin{minipage}[m]{0.6\textwidth}
     \includegraphics[width=1.2\linewidth]{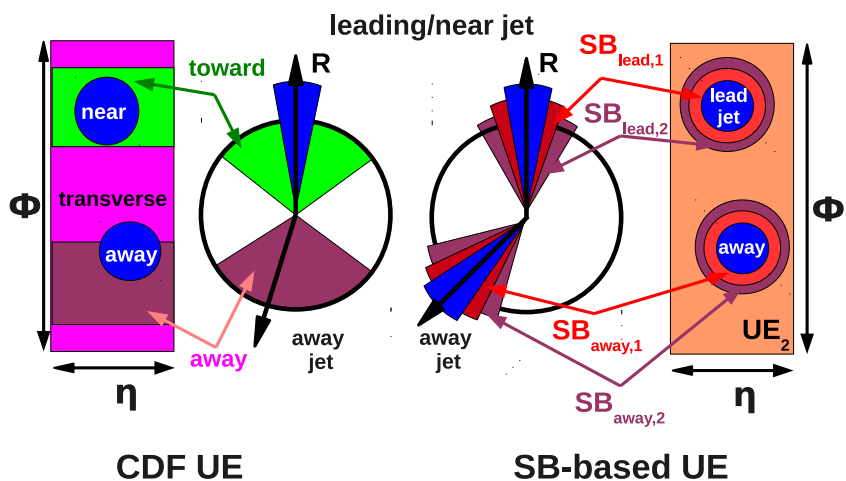}
  \end{minipage}
\caption{({\it Color on-line.}) The schematic view of the underlying
event (UE) defined by the CDF-method (left) and the SB-method (right). Details
can be found in Ref.~\cite{hcbm:2010}.}
\label{fig:cdf-sb-ue}
\end{figure*}

We introduced a new definition in 
Refs.~\cite{Agocs:2009,Agocs:2010,bggmex:2010,hcbm:2010} in agreement 
with the layout of Fig.~\ref{fig:cdf-sb-ue}, where the left side   
displays the CDF-based definition, and SB-based UE is plotted on the right side. 
As it can be seen on Fig.~\ref{fig:cdf-sb-ue}, the main difference between 
the two definitions is the multiple application of jet identification 
with jet-cone angle $R=\sqrt{\Delta \Phi^2 + \Delta \eta^2}$, 
and setting an approximate dial-like area, around 
which concentric bands (or rings) are the surrounding belts, 
denoted by `$SB_1$' and `$SB_2$'. 
The thickness of $\delta R_{SB1}=\delta R_{SB2}$ is $0.1$.

\section{The Analysis of Underlying Event Definitions}

One of the aims of SB-based definition is to separate the jet-like 
particles from the soft (or semi-hard) ones. This can be carried out via 
a comparison of physical quantities e.g.: (i) the average number of hadrons  
within the defined areas relative to the total event multiplicities, 
and (ii) the transverse momentum distributions of hadrons in the discussed regions. 
Below we recall the original CDF-based and our new SB-based definitions 
in order to test the validity of the SB-based definition.  

We performed an extended study of an SB-based analysis of a simulated
data set for $pp$ collisions at 7 TeV -- similar to which was
published in Refs.~\cite{bggmex:2010,hcbm:2010}.  We have analyzed a
data set of 739 500 $pp$ events generated by PYTHIA~6~\cite{pythia62},
Perugia-0~tune~\cite{Perugia0}.  This sample is similar to the
LHC10e14 sample calculated within the ALICE framework. The sample was
generated with constrained $p_{T,hard}$ interval
$[10\,GeV/c,20\,GeV/c]$ Jets are identified by the UA1
method~\cite{UA1}, setting $R=0.4$. Applying the cuts we got 174 452
two-jet events. For summing up hadron numbers correctly within the
selected specific regions, proper determination of the areas is
needed. In parallel, selected jets identified in the ALICE TPC's
acceptance ($|\eta|<0.9$) must be counted properly as well.

\begin{figure*}[h]
  \centering
  \begin{minipage}[l]{0.46\textwidth}
    \includegraphics[width=\linewidth]{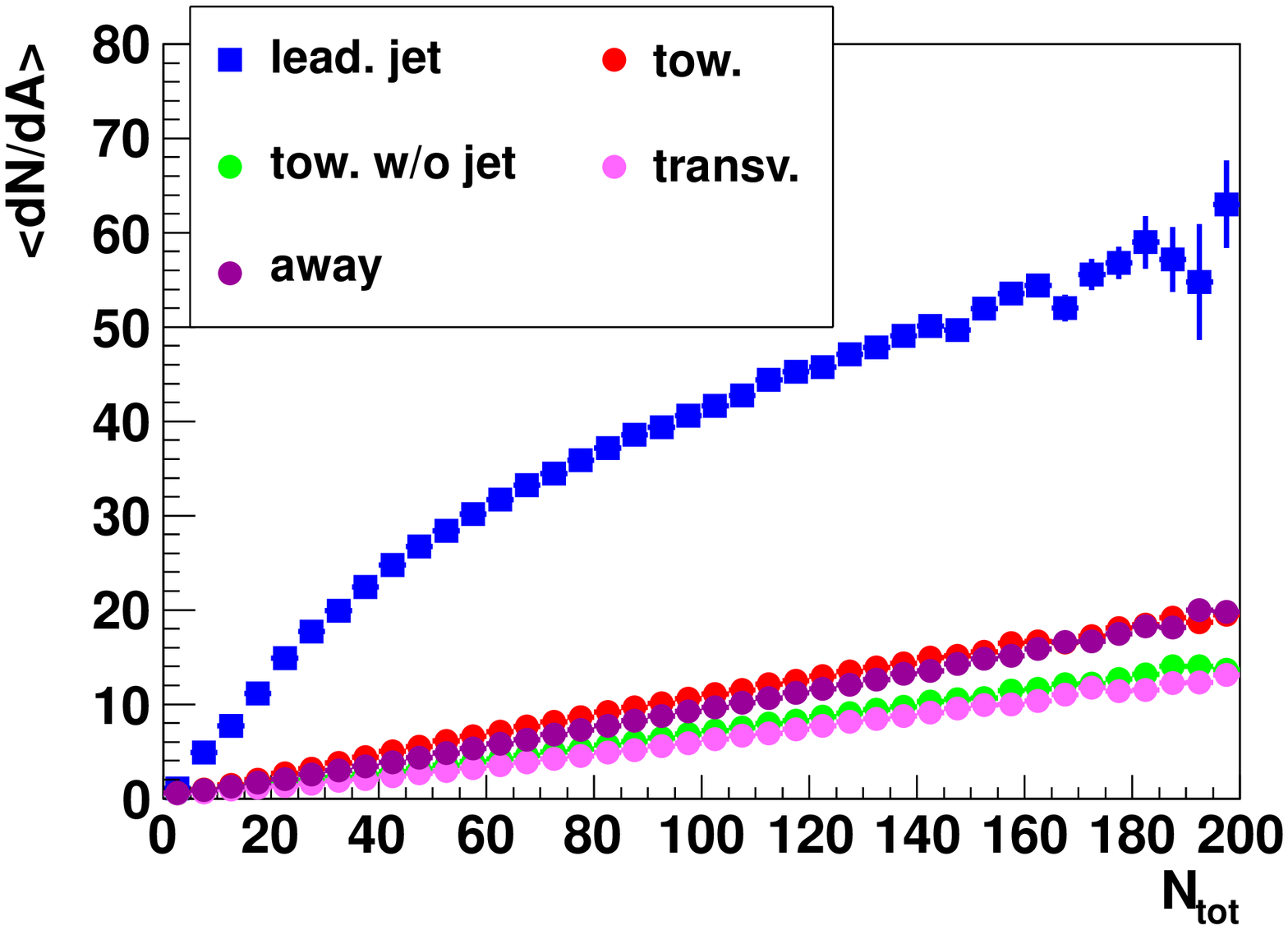}
  \end{minipage}
  \begin{minipage}[r]{0.46\textwidth}
    \includegraphics[width=\linewidth]{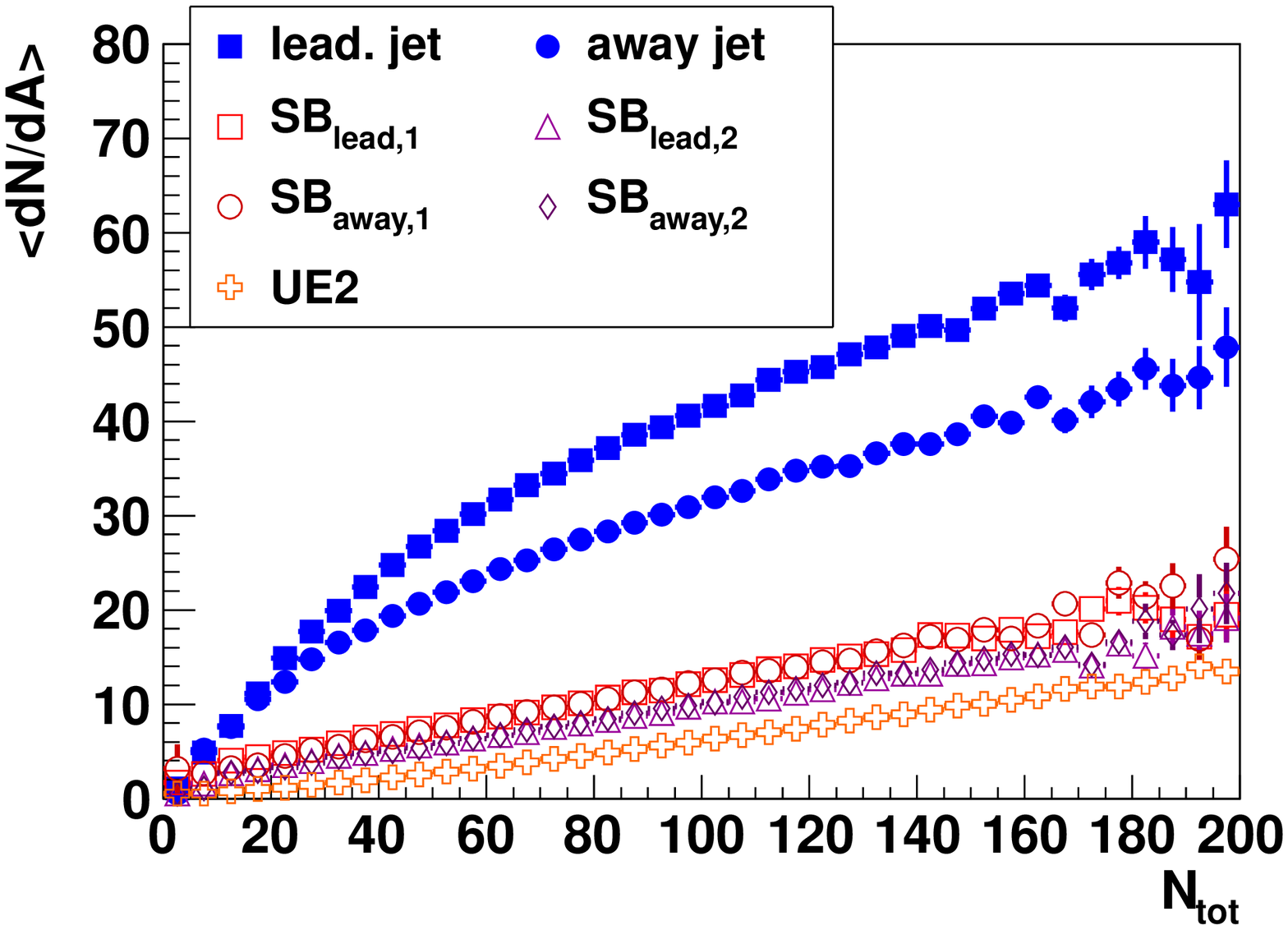}
  \end{minipage}
  \begin{minipage}[l]{0.46\textwidth}
    \includegraphics[width=\linewidth]{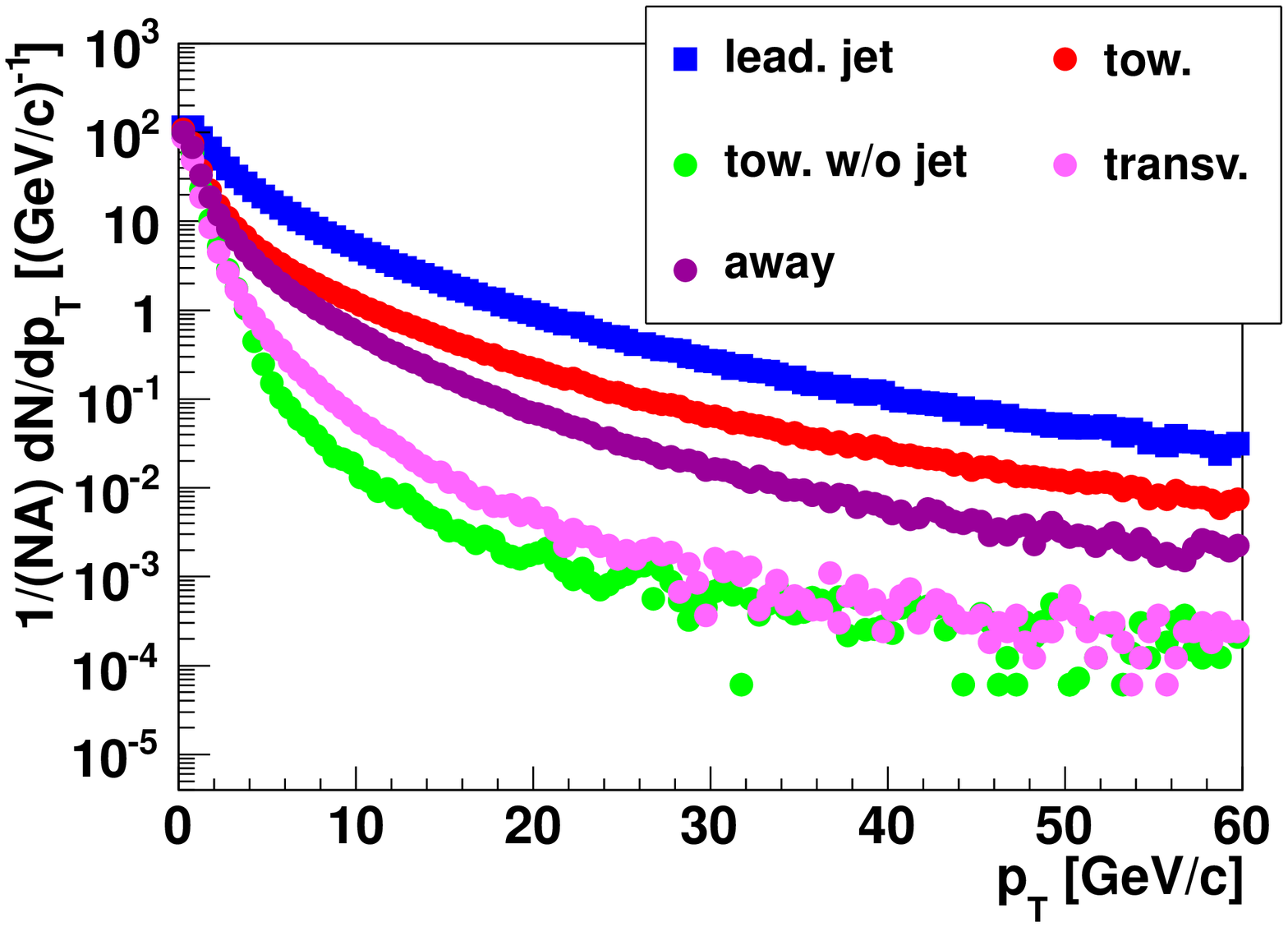}
  \end{minipage}
  \begin{minipage}[r]{0.46\textwidth}
    \includegraphics[width=\linewidth]{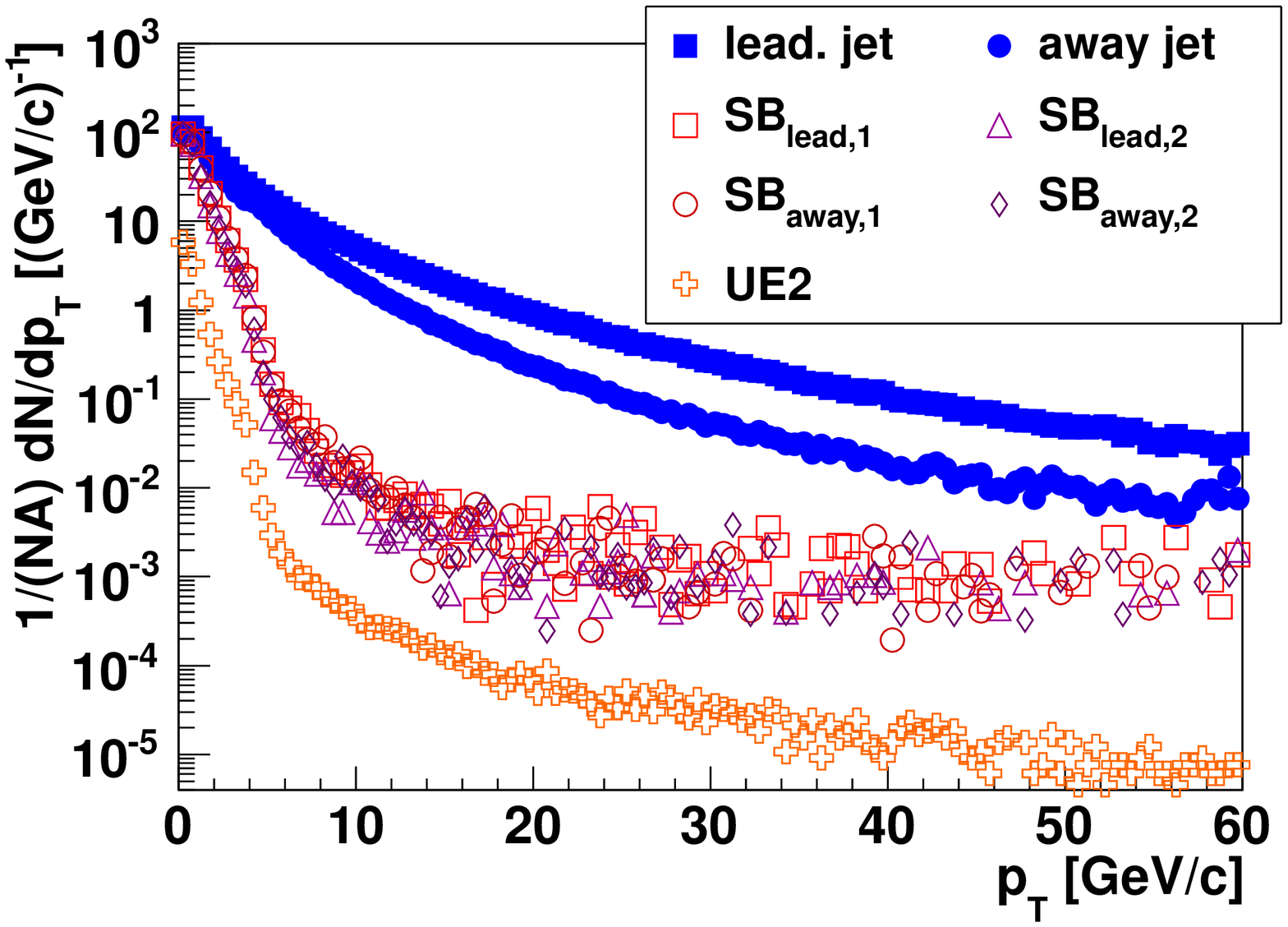}  
  \end{minipage} 
\caption{({\it Color on-line.}) The area averaged charged hadron 
multiplicities, $d N/d A$, of the selected areas depending on the total 
multiplicities of the events, $N_{tot}$ ({\it upper panels}) and
transverse momentum distribution of selected hadrons ({\it lower panels}) 
for CDF-based ({\it left panels}) and for SB-based ({\it right panels}) results. }
\label{fig:n-vs-ntot}
\end{figure*}

\section{Results and Conclusions}

Fig.~\ref{fig:n-vs-ntot} compares the CDF-based ({\it left}) 
and SB-based ({\it right}) definitions applied on $pp$ collisions 
at 7 TeV center of mass energy. {\it Upper panels} present 
the area averaged charged hadron multiplicities, $d N/d A$ vs. 
the number of total multiplicity, $N_{tot}$. 
{\it Lower panels} display the $p_T$ distribution of selected hadrons 
in the given regions. 

{\it Upper panels} of Fig.~\ref{fig:n-vs-ntot} show that area averaged
charged hadron multiplicities are quite similar in both definitions. 
The jet-content areas ({\it blue full squares for near and disks for away sides}) 
clearly characterized by larger average values. Away side ({\it purple full dots}), 
surrounding belts ({\it open circles and triangles}), 
and underlying event ({\it pink dots for CDF and orange open crosses for SB}) 
give similar values in both cases in decreasing order. 
In both cases the UE has the smallest value proportional to the $N_{tot}$. 
In this way the two definitions give the same result.

The $p_T$ distributions of fluxes for the selected areas are given 
on the {\it lower panels} of Fig.~\ref{fig:n-vs-ntot} using the same 
notations and colors as above. Similarly to the {\it upper panels} 
the distributions are higher for the jet-content areas and getting smaller 
to the direction of away side, surrounding belts, and underlying event. 
It is interesting to see, that for the SB-based case all spectra are almost the same, 
but the separation of jet-content, SBs, and UE are well defined. 
Moreover, the shapes of the curves clearly reflect the origin of the hadrons: 
jet-content distributions are power-law like, 
but the SBs and UE areas are exponential at lower $p_T$s, 
indicating the bulk origin of the hadrons found in these areas. 

In summary, the SB-method, especially the $p_T$ distributions of fluxes,
gives good separation of the charged hadron yields from 
different regions and its general use is supporting the of study 
the properties of UE and any jet-matter interactions inside the SBs.

\section*{Acknowledgments}
This work was supported by Hungarian OTKA NK77816, PD73596 and
E\"otv\"os University. Author GGB thanks for the J\'anos Bolyai 
Research Scholarship of the HAS.


\end{document}